\documentclass[journal=aanmf6,manuscript=article,layout=twocolumn]{achemso}

\usepackage[T1]{fontenc} 
\usepackage{graphicx}
\usepackage{amsmath}
\usepackage{float}
\usepackage[colorlinks=true, urlcolor=blue, citecolor=blue, linkcolor=blue]{hyperref}
\usepackage{titlesec}
\titleformat*{\section}{\bfseries \sffamily}

\author{\flushleft \textrm{Bernd Aichner}}
\affiliation{Faculty of Physics, University of Vienna, Wien, Austria}
\author{\textrm{Benedikt M\"uller}}
\author{\textrm{Max Karrer}}
\affiliation{Physikalisches Institut and Center for Quantum Science (CQ) in LISA$^+$, Universit\"at T\"ubingen, T\"ubingen, Germany}
\author{\textrm{Vyacheslav R. Misko}}
\affiliation{Department of Physics, Universiteit Antwerpen, Antwerpen, Belgium}
\alsoaffiliation{Theoretical Quantum Physics Laboratory, RIKEN Cluster for Pioneering Research, Wako-shi, Saitama, Japan}
\alsoaffiliation{$\mu$Flow group, Department of Chemical Engineering, Vrije Universiteit Brussel, Brussels, Belgium}
\author{\textrm{Fabienne Limberger}}
\affiliation{Physikalisches Institut and Center for Quantum Science (CQ) in LISA$^+$, Universit\"at T\"ubingen, T\"ubingen, Germany}
\author{\textrm{Kristijan L. Mletschnig}}
\affiliation{Faculty of Physics, University of Vienna, Wien, Austria}
\author{\textrm{Meirzhan Dosmailov}}
\affiliation{Institute of Applied Physics, Johannes Kepler University Linz, Linz, Austria}
\author{\textrm{Johannes D. Pedarnig}}
\affiliation{Institute of Applied Physics, Johannes Kepler University Linz, Linz, Austria}
\author{\textrm{Franco Nori}}
\affiliation{Theoretical Quantum Physics Laboratory, RIKEN Cluster for Pioneering Research, Wako-shi, Saitama, Japan}
\alsoaffiliation{Physics Department, University of Michigan, Ann Arbor, USA}
\author{\textrm{Reinhold Kleiner}}
\affiliation{Physikalisches Institut and Center for Quantum Science (CQ) in LISA$^+$, Universit\"at T\"ubingen, T\"ubingen, Germany}
\author{\textrm{Dieter Koelle}}
\affiliation{Physikalisches Institut and Center for Quantum Science (CQ) in LISA$^+$, Universit\"at T\"ubingen, T\"ubingen, Germany}
\author{\textrm{Wolfgang Lang}}
\email{wolfgang.lang@univie.ac.at}
\affiliation{Faculty of Physics, University of Vienna, Wien, Austria}

\title[Ultradense Pinning Arrays in YBCO]
  {{\small \sffamily This is an author-created version of an article published in\\ ACS Applied Nanomaterials, {\bf 2}, 5108--5115 (2019)\\ DOI: 10.1021/acsanm.9b01006. \textcopyright 2019 American Chemical Society.\\  The Version of Record is available online at \href{https://doi.org/10.1021/acsanm.9b01006}{https://doi.org/10.1021/acsanm.9b01006}
  .} \vskip 10 pt \hrule \vskip 20pt
  \flushleft \Large \bf \sffamily Ultradense Tailored Vortex Pinning Arrays in Superconducting YBa$_2$Cu$_3$O$_{7-\delta}$ Thin Films Created by Focused He Ion-Beam Irradiation for Fluxonics Applications}

\begin{document}

\onecolumn
\hrule
\begin{center}
\includegraphics[width=12 cm]{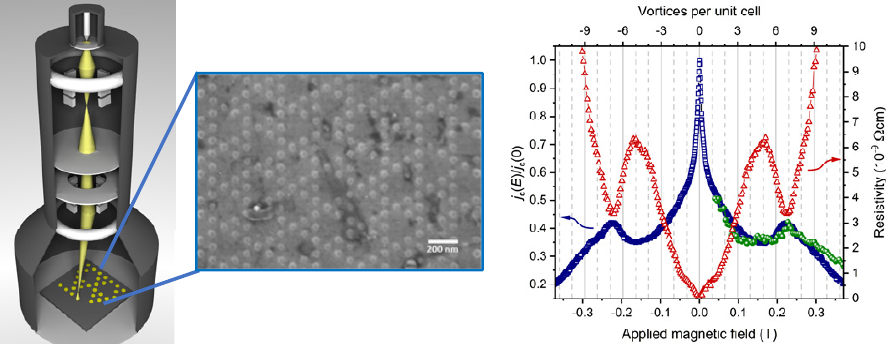}\\
\end{center}
{\bf ABSTRACT:} Magnetic fields penetrate a type II superconductor as magnetic flux quanta, called vortices. In a clean superconductor they arrange in a hexagonal lattice, while by addition of periodic artificial pinning centers many other arrangements can be realized. Using the focused beam of a helium ion microscope we have fabricated periodic patterns of dense pinning centers with spacings as small as 70\,nm in thin films of the cuprate superconductor YBa$_{2}$Cu$_{3}$O$_{7-\delta}$. In these ultradense kagom\'e-like patterns, the voids lead to magnetic caging of vortices, resulting in unconventional commensurability effects that manifest themselves as peaks in the critical current and minima in the resistance versus applied magnetic field up to $\sim 0.4\,$T. The various vortex patterns at different magnetic fields are analyzed by molecular-dynamics simulations of vortex motion, and the magnetic field dependence of the critical current is confirmed. These findings open the way for a controlled manipulation of vortices in cuprate superconductors by artificial sub-100\,nm pinning landscapes.
\vskip 5pt
\hrule
\twocolumn

\section{INTRODUCTION}
Superconductivity is one of the most intriguing phenomena in condensed matter physics and in particular the
cuprate superconductors pose a challenge to the understanding of their electron-pairing mechanism. Still, their high transition temperature $T_c$ and their possible high operating temperature in the accessible range of reliable and easy-to-handle cryocooler technology, make the cuprate superconductors primary candidates for emerging technologies. However, all superconductors are only marginally suitable for technical applications in their pure and clean form. It is by the introduction of controlled defects that superconductors can be tailored for many important properties, e.g., by enhancing their ability to carry a lossless current, which requires to impede the dissipative motion of magnetic flux quanta, also called Abrikosov vortices or fluxons \cite{KLEI16M}.

Whereas such enhanced vortex pinning in con\-ventional metallic superconductors, typical\-ly used for medical applications and high-field magnets, has been achieved by metallurgical techniques, the brittle nature of cuprate high-$T_c$ superconductors (HTSCs) requires different concepts. A successful approach is to create columnar amorphous regions with diameters of a few times the superconducting coherence length by irradiation with swift heavy ions \cite{CIVA97R}.

In contrast to these extended defects, point defects can be created by inclusion of tiny nonsuperconducting particles \cite{HAUG04}, or by electron, proton, and light ion irradiation of HTSCs. For energies up to few megaelectronvolts the incident particles collide with a nucleus and displace it, eventually creating a collision cascade for high enough recoil energies. Several studies \cite{SEFR01,LANG04,CYBA14a} have revealed that irradiation with He$^+$ ions of moderate energy is well suitable to tailor the superconducting properties in thin films of the prototypical HTSC YBa$_{2}$Cu$_{3}$O$_{7-\delta}$ (YBCO) by displacing mainly oxygen atoms, leading to a controllable reduction or even complete suppression of $T_c$. Arrays of cylindrical defect channels (CDs) that are populated with point defects provide a landscape in which superconductivity is locally suppressed. Recently, it has been demonstrated \cite{LANG06a,LANG09,PEDA10,SWIE12,TRAS13,HAAG14,ZECH17a,YANG19P} that such pinning potential landscapes allow one to accommodate vortices in a commensurate arrangement, leading to peaks in the critical current and minima of the resistance at well-defined matching magnetic fields.

In the field of vortex commensurability effects, a large body of research has been established for metallic superconductors, typically using an array of holes (antidots) or blind holes with about $1\,\mu$m spacings \cite{MOSH11M}. Considerably narrower spacings have been achieved for thin Nb films grown on porous substrates \cite{WELP02,HALL09}. However, in these superconductors, the Ginzburg--Landau coherence length at zero temperature is on the order of $\xi(0) \sim 10$~nm and the London penetration depth on the order of $\lambda_L(0) \sim 100$~nm, the latter setting the range of magnetic interaction between vortices. Commensurability effects could be explored only at temperatures $T$ very close to $T_c$, taking advantage of the fact that $\lambda_L(T)=\lambda_L(0)[T_c/(T_c-T)]^{1/2}$ increases substantially when approaching $T_c$. But choosing a rather high operation temperature is not feasible for practical applications since it is on the expense of a reduced superconducting gap, weaker pinning potential, and enhanced thermodynamic fluctuations.

On the other hand, thin films of YBCO have an in-plane $\lambda_{L,ab}(0) \sim 250$~nm \cite{WOLB14,THIE16,MART17,ROHN18}, and edge and screw dislocations, as one important type of intrinsic defects in YBCO films, have typical distances of about 300~nm \cite{DAM99}. Still, commensurability effects could be demonstrated close to $T_c$ in YBCO perforated with a square array of holes with $1\,\mu$m lattice spacing \cite{CAST97}. But since intrinsic defects compete with the artificially created pinning landscape, it is important to fabricate and investigate pinning arrays with lattice spacings significantly below these two characteristic lengths. Here, we report on the electrical transport properties of ultradense unconventional pinning arrays in YBCO thin films with spacings down to 70\,nm, fabricated with a helium ion microscope, and we demonstrate pronounced commensurability effects at strong magnetic fields well above 100\,mT which persist down to at least $\sim 50\,$K, i.e., far below $T_c$.

\section{RESULTS AND DISCUSSION}

\begin{figure*}[t!]
\includegraphics[width=\textwidth]{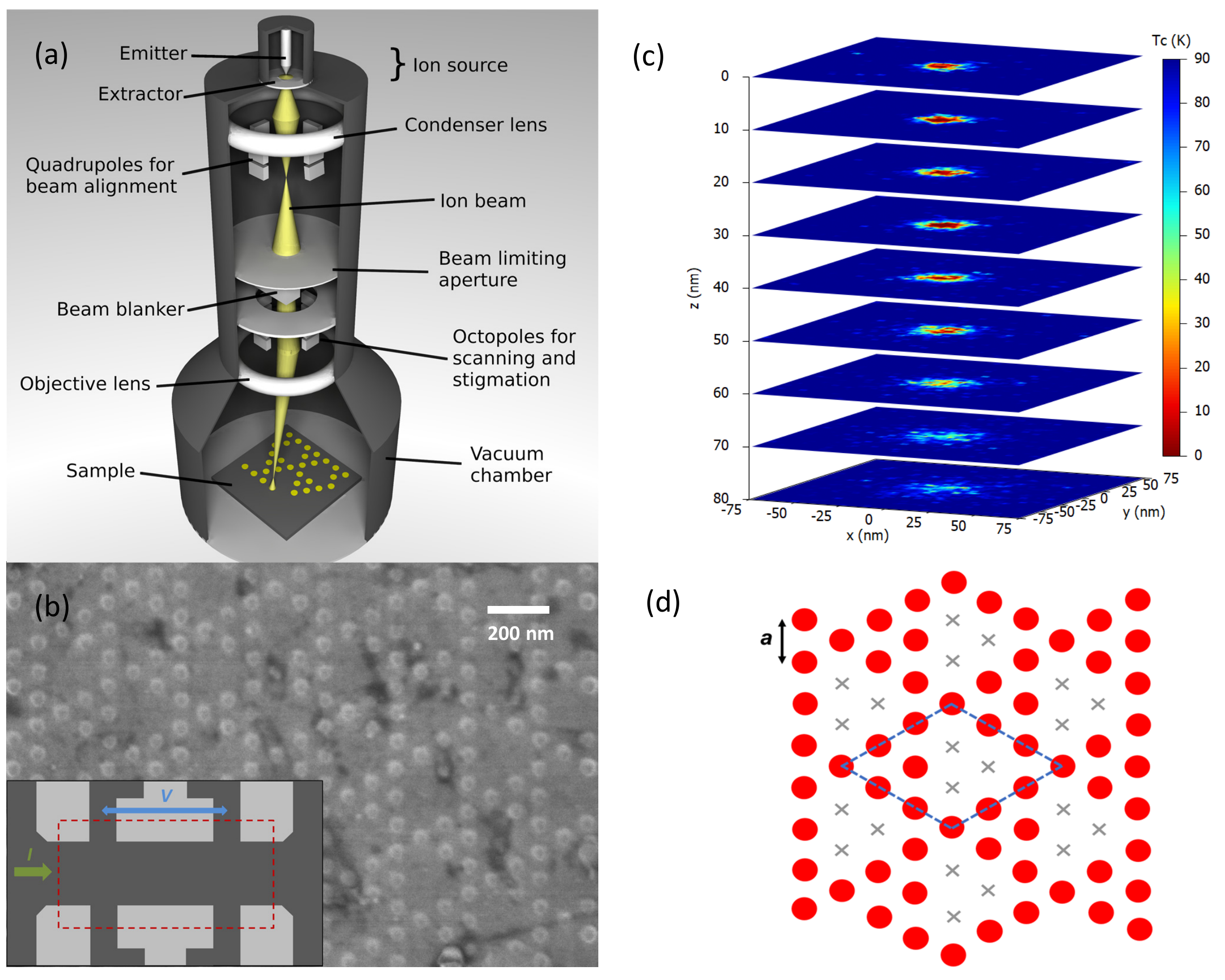}
\caption{
(a) Schematic drawing of direct modification of the local superconducting properties in a HIM.
(b) HIM image of a quasi-kagom\'e test pattern of HIM-induced defects in a YBCO film, irradiated with a fully focused beam and with $5\times10^6$ ions/spot, i.e., with a 333 times larger number than in (c). The lattice spacing is $a=90\,$nm. Inset: sketch of the YBCO thin film samples (dark gray), the $200\,\mu \text{m} \times 100\,\mu \text{m}$ rectangular irradiated area (red broken lines), direction of the applied current $I$, and the voltage probes $V$.
(c) Simulation of defect-induced spatial distribution of $T_c$ of a YBCO film (within sheets at various depths $z$ from the surface $z=0$), upon irradiating one spot with a defocused 30~keV He$^+$ ion beam with a  Gaussian normal distributed fluence of 20~nm fwhm diameter. The number of impacting ions (15000) is the same as in the experiments (per spot) and corresponds to an average fluence of $4.8 \times 10^{15}\,\text{cm}^{-2}$.
(d) Sketch of the pinning lattice with lattice spacing $a$: red disks represent irradiated spots, the blue dashed lines indicate the unit cell of the quasi-kagom\'e tiling, and gray crosses mark those sites that were removed from the hexagonal lattice to form the quasi-kagom\'e tiling.}
\hrule
\label{fig:him_all}
\end{figure*}

{\bf \sffamily 2.1 Irradiation in the Helium Ion Microscope.\ }
The ZEISS ORION NanoFab \cite{ORION} is a novel multifunctional platform combining focused ion beam sources for neon and helium ions and a scanning helium ion microscope (HIM) with a spatial resolution better than 0.5~nm and unprecedented depth of focus \cite{WARD06,HLAW16M,FLAT18}. The HIM consists of a gas-field He$^+$ ion source that emits ions from an ultrasharp tip, electrostatic ion optics to focus and trim the beam, and a deflection system to raster the beam over the sample stage as it is outlined in Figure~\ref{fig:him_all}a. The advantage of using a He beam over the conventional Ga focused ion beam technique is its higher spatial resolution (fabrication of nanopores down to 1.3~nm has been demonstrated \cite{EMMR16}) and the prevention of contamination of the HTSC by Ga ions. As a first application in YBCO, the fabrication of Josephson junctions and superconducting quantum interference devices has been demonstrated via forming thin barriers of insulating material with the focused ion beam across prepatterned microbridges \cite{CYBA15,CHO15,CHO18,MULL19}.

However, for the fabrication of vortex pinning defects, the HIM in its original operation mode is less suitable. On the one hand, it is known that the diameter of defects suitable for vortex pinning must not be smaller than the superconducting coherence length, which is $\xi_{ab}(0) \simeq 1.2\,$nm in YBCO \cite{SEKI95}. On the other hand, the inherent deflection of the ion trajectories by scattering at the target atoms when traversing the material will cause a blurring of the focus with increasing depth. In this situation the fluence of the He$^+$ ions decreases strongly from the surface to the back of the YBCO film. To achieve a high enough ion fluence for suppression of $T_c$ throughout the entire film thickness, the intensity of the focused ion beam must then be set to such a high value that amorphization of the YBCO crystal lattice \cite{LANG06a,MULL19} and mechanical destruction near the sample's surface result. This is illustrated in Figure~\ref{fig:him_all}b which was recorded in the HIM after irradiation of a test pattern with an optimally focused He$^+$ ion beam with a much larger number of ions/spot than normally used in our experiments. Around the impact point of the focused ion beam the YBCO structure is completely destroyed and blisters with about 50~nm diameter bulge out of the surface, while under the regular conditions for fabrication of pinning lattices, the irradiated areas are invisible in both the HIM and with scanning electron microscopy.

To avoid these shortcomings and irradiate a well-defined area, our approach uses an intentionally defocused ion beam. This is achieved by first adjusting the HIM settings to highest resolution and afterward changing the working distance (beam focus plane) by $8\,\mu$m. The aperture angle of the ion beam is  $\pm 0.07^\circ$; hence, the ion beam hits the sample surface almost orthogonally with a nearly Gaussian fluence profile \cite{EMMR16} with a full width at half maximum (fwhm) of about 20~nm.

The resulting three-dimensional shape of CDs with suppressed $T_c$ is inferred from simulations of ion--matter interaction and collision cascades with the program SRIM \cite{ZIEG10} and calibrated to experimental data of $T_c$ suppression as a function of defect density. Details of such calculations are reported elsewhere \cite{MLET19}. A defocused 30~keV He$^+$ beam of 20~nm fwhm diameter creates well-defined cylindrical channels with diameters $\sim 25\,$nm in a YBCO film within which $T_c$ is suppressed or reduced up to a depth of 80~nm, as evidenced in Figure~\ref{fig:him_all}c. In thicker YBCO films, however, the collision cascades are prone to substantial straggling which leads to blurring of the profile and insufficient reduction of $T_c$ due to a lower defect density as it has been observed in heavy--ion--irradiated YBCO single crystals \cite{NIEB01}.

{\bf \sffamily 2.2 Vortex Commensurability Effects in Quasi-Kagom\'e Vortex Pinning Arrays.\ }
As a proof-of-concept experiment we have fabricated a square array of columnar defect cylinders with a defocused ion beam of fwhm\,=\,50\,nm and with a lattice spacing (nearest--neighbor distance) $a=200\,$nm in a 80\,nm thick YBCO film and measured the critical current $I_c(B_a)$ in the superconducting state and the magnetoresistance $R(B_a)$ in the voltage state as a function of the applied magnetic field $B_a$. The parameters were chosen similarly to previous experiments performed with masked ion beam irradiation \cite{SWIE12,TRAS13,HAAG14,ZECH17a}. Consistently, a clear peak of $I_c(B_a)$ corresponding to a minimum of $R(B_a)$ is found when each CD is filled by one vortex, and a tiny feature can be noticed for filling of each CD by two vortices (see the Supporting Information).

Pinning lattices can be designed not only with common hexagonal \cite{OOI05} or square tilings \cite{MOSH11M} but also as more complex periodic or even quasi-periodic tilings that exhibit a number of unusual phenomena. Some examples of such arrays  have been studied theoretically \cite{LAGU01,MISK05,MISK06,REIC07a} and experimentally in metallic superconductors with holes \cite{KEMM06,MISK10,BOTH14} and magnetic dots \cite{VILL06,SILH06,KRAM09,LARA10}, like Penrose \cite{KEMM06,MISK10}, honeycomb \cite{WELP02,LATI12} and kagom\'e \cite{CUPP11} lattices and artificial vortex ice arrangements in geometrically frustrated pinning lattices \cite{LIBA09,LATI13,XUE18}. In HTSCs, such studies have been scarce \cite{TRAS14} due to the much more demanding nanopatterning.

\begin{figure*}[t!]
\includegraphics[width=\textwidth]{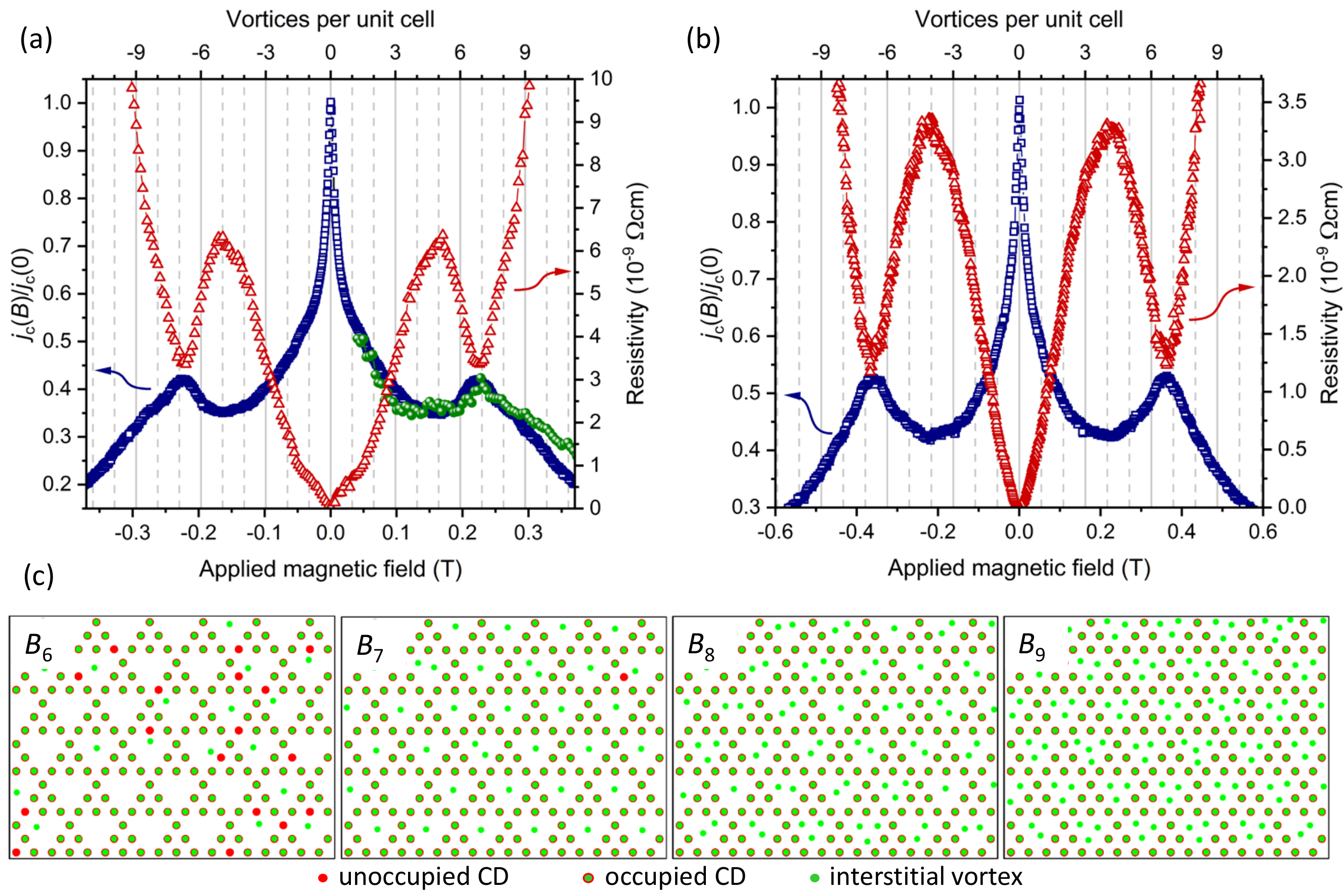}
\caption{
(a) Magnetic field dependence of the normalized critical current at 83\,K (blue squares) and resistivity at 83.5\,K (red triangles) of a 75\,nm thick YBCO film, irradiated with a quasi-kagom\'e pattern of beam spots (20~nm fwhm) in a HIM (sample QK90). The upper horizontal scale indicates the matching fields $B_k$, calculated from eq~\ref{eq:matching}, for a filling of the unit cell with $k$ vortices. The peak in the critical current and the minimum in the resistivity correspond to $k = 7$. The green bullets represent the results from molecular dynamics simulations. Only the positive branch of the mirror symmetric results is shown for clarity. (b) Normalized critical current at 52\,K (blue squares) and resistivity at 55\,K (red triangles) for sample QK70. (c) Simulation of vortex arrangements in the quasi-kagom\'e lattice of sample QK90 for several applied magnetic fields that correspond to various commensurability conditions as labeled in the graphs. Red circles represent artificial pinning centers and green dots the vortices.}
\hrule
\label{fig:match_all}
\end{figure*}

The particular advantage of irradiation in the HIM over other nanopatterning techniques is the higher spatial resolution and the ability to easily produce any desired pattern. We demonstrate this by the fabrication of quasi-kagom\'e pinning lattices with ultrasmall lattice spacing $a=90$\,nm (sample QK90) and 70\,nm (sample QK70). The quasi-kagom\'e lattice is formed from a hexagonal lattice, where three neighboring sites are not occupied \cite{LARA10}, as sketched in Figure~\ref{fig:him_all}d. The  quasi-kagom\'e lattice consists of vertex triangles with six sites per unit cell and large voids. It provides a unique platform to investigate the competition between pinning potentials and elastic energy of the vortex lattice. The quasi-kagom\'e lattice is a more extreme variant of the genuine kagom\'e tiling, which is formed from the hexagonal lattice by eliminating every other site from every other row.

In periodic pinning lattices commensurability effects are expected to occur at applied magnetic fields that fulfill the matching condition
\begin{equation}
\label{eq:matching}
B_k = \frac{k \Phi _0}{A}
\end{equation}
where $k$ is the number of pinning sites (or vortices) in the unit cell of area $A$ and $\Phi_0=h/(2e)$ is the magnetic flux quantum.

The electrical transport properties of the YBCO films patterned with the quasi-kagom\'e pinning lattice show intriguing features. Those can be attributed to the fact that HIM nanopatterning enables the realization of ultradense pinning arrays. Accordingly, pronounced vortex commensurability effects in the critical current density $j_c(B_a)$ and the magnetoresistivity $\rho(B_a)$ appear at very high matching fields, at $B_a = 0.23\,$T for sample QK90 and at $B_a = 0.38\,$T for sample QK70, as shown in Figures~\ref{fig:match_all}a and \ref{fig:match_all}b, respectively. The observed effects are well reproducible after storing the samples at room temperature for several weeks.

In the case of strong interactions between vortices and the pinning potential of the CDs, significant vortex commensurability effects are expected at $B_6$, when each CD is filled with one vortex. Conversely, with strong vortex--vortex interactions and  marginal influence of the pinning landscape, the elastic energy of the vortex ensemble should favor a hexagonal lattice and the voids are expected to be filled with three interstitial vortices, leading to commensurability signatures at $B_9$. The former, for instance, was observed in a quasi-kagom\'e lattice of magnetic Ni dots embedded in a Nb superconducting film \cite{LARA10}.

Surprisingly, none of these features are found here; instead, a prominent matching peak in $j_c(B_a)$ and a corresponding minimum in $\rho(B_a)$ appear at $B_7$, as can be seen in Figure~\ref{fig:match_all}a,b. Presumably, all CDs are occupied by vortices and one interstitial vortex per unit cell is ``caged'' inside the voids of the lattice due to repulsion forces from the trapped vortices. On the other hand, the absence of a matching feature at $B_1$ indicates that the six closely packed sites of the kagom\'e triangle do not merge into a single pinning site. This confirms that despite the rather narrow distance of CDs they act as separate pinning potentials.

Molecular dynamics simulations shed more light on the vortex distribution in the quasi-kagom\'e lattice.
A rectangular cell with periodic boundary conditions is used for the calculations. The vortex--vortex interaction, the pinning force on the vortices at the CDs, an external driving force, and a thermal stochastic force are taken into account, as detailed elsewhere for previous simulations \cite{REIC98b,MISK05,MISK06,MISK12}. The ground state of the static vortex arrangement (zero driving force) is obtained by starting the simulation at some elevated ``temperature'' and gradually decreasing it to zero, thus performing a simulated thermal annealing. Note that this corresponds to magnetic-field-cooled experiments. However, since in our measurements no hysteresis was observed between up- and down-ramping of the magnetic field, these results are equally valid for magnetic-field-ramped experiments.

In the simulations, the critical current density $j_c$ is determined to be the minimum current density that depins the vortices; in the experiments the common electric field criterion of $10\,\mu$V/cm is used. The simulated values for the critical current density are then scaled to the experimental results (for sample QK90) at $B_a=50\,$mT.

The results of the simulations of static vortex configurations at different magnetic fields $B_k$ and for temperatures adapted to the experimental situation are displayed in Figure~\ref{fig:match_all}c.
For $B_6$, where the average density of vortices matches the density of CDs, most of the pinning centers are filled with vortices, but several empty pins and, in turn, interstitial vortices can be noticed. This is explained by the strong repulsion between the vortices located on the closely packed pinning sites such that some of these vortices can depin and move in the voids thus minimizing the energy. For $B_7$ almost all pins are filled and typically one excess vortex sits in the centers of the voids of the quasi-kagom\'e lattice. If the magnetic field is further increased, all pins are occupied and the voids host two ($B_8$) or three ($B_9$) vortices with minor fluctuations $\pm 1$ in this number. In the latter case, the hexagonal vortex lattice is mostly recovered. Note that an experimental visualization of the vortex arrangements in ultradense pinning lattices is challenging. Because $\lambda_{L,ab}(T) \gg a$ in our samples, the magnetic fields of the vortices overlap strongly and magnetic force microscopy might not provide enough contrast.

\begin{figure*}[t!]
\includegraphics[width=\textwidth]{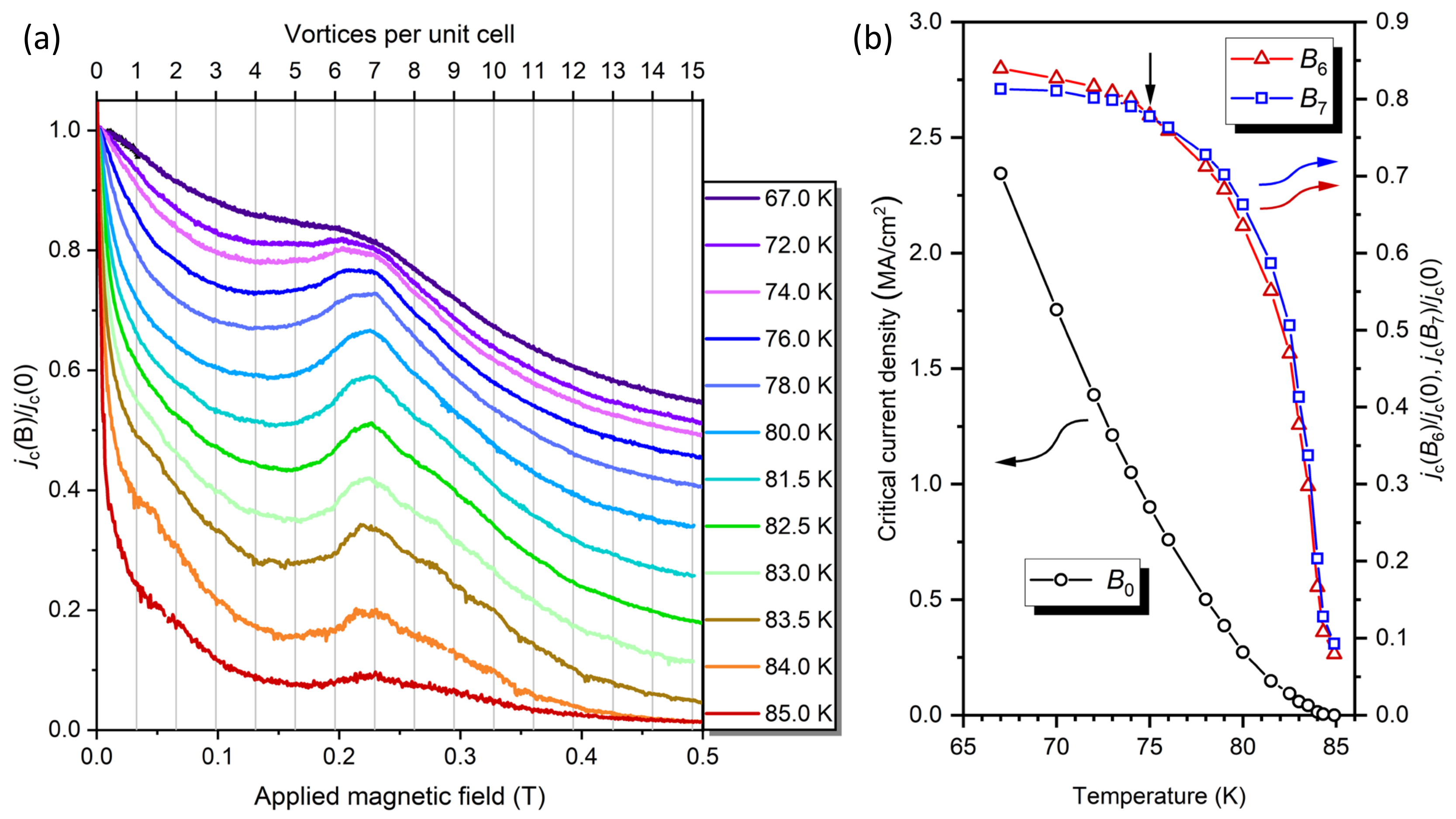}
\caption{
(a) Normalized critical current density of sample QK90 as a function of the applied magnetic field at various temperatures. The branches at reversed polarity of the magnetic field are mirror symmetric and are not shown. The upper horizontal scale displays the commensurability fields for $k$ vortices in the unit cell of the quasi-kagom\'e pattern.
(b)~Temperature variation of the critical current density at $B_a=0$ (black circles) and the normalized critical current densities at $B_6$ (red triangles) and $B_7$ (blue squares). At lower temperatures, starting from the crossover at 75\,K marked by the arrow, the critical current at the $B_6$ commensurability field exceeds the critical current at $B_7$.}
\label{fig:Jc_T}
\hrule
\end{figure*}

A simulation of the critical current density $j_c(B_a)$ is displayed in Figure~\ref{fig:match_all}a. Indeed, there is no peak in $j_c(B_a)$ visible at $B_6$, and peaks of $j_c(B_a)$ at $B_8$ and $B_9$ are hardly noticeable, whereas a pronounced peak at $B_7$ is found, in excellent agreement with the experimental results. This unconventional commensurability effect can be explained by the competition of the elastic vortex lattice energy, which favors a hexagonal lattice and the semiregular quasi-kagom\'e pinning potential. At $B_6$ filling of all CDs with vortices would lead to a maximized pinning force density, but the repulsion between the trapped vortices leads to a significant number of vacancies at the pinning sites and, in turn, mobile interstitial vortices in the kagom\'e voids that reduce $j_c$. On the other hand, for $B_9$, the elastic energy of the vortex lattice is minimized, but pinning is not so efficient due to the larger number of interstitial vortices. Thus, a commensurate arrangement, in which all pinning sites are occupied and one interstitial vortex is caged in the center of the voids appears to be the most stable structure at temperatures near $T_c$.

The balance between the pinning forces at the CDs and the vortex caging potential can be tuned with temperature. While both increase at lower temperature, the pinning forces at the artificial defect lattice increase more rapidly \cite{KHAL93}. To study this behavior on our samples, we performed electric transport measurements at variable $T$. Figure \ref{fig:Jc_T}a shows normalized $j_c$ vs $B_a$ curves for sample QK90 at $67\,{\rm K}\le T \le 85\,$K. We find that at temperatures $T < 80\,$K a second peak of $j_c$ around $B_6$ emerges, coalescing with the peak at $B_7$ into a broader feature with $j_c(B_6) > j_c(B_7)$ at even lower temperatures. This is a formidable demonstration of how the vortices ``crystallize'' into the quasi-kagom\'e lattice at lower temperatures when the artificial pinning landscape becomes dominant. Figure \ref{fig:Jc_T}b displays the critical current density $j_c$ at $B=0$, calculated from the sample's cross section. However, it has to be cautioned that the irradiated sample contains many CDs, and hence a significant volume in the sample does not contribute to the supercurrent, making a comparison to pristine YBCO films inappropriate. The normalized $j_c(B_6)/j_c(0)$ and $j_c(B_7)/j_c(0)$ curves reveal the crossover at $T=75\,$K, where pinning at a matching field of six vortices/unit cell becomes more efficient.

Our findings may be compared to somewhat related simulation results. For a honeycomb pinning lattice it has been demonstrated \cite{REIC07a} that $j_c$ of a commensurable vortex arrangement with an additional caged vortex can be larger than that with the pinning sites occupied only. However, this only applies to shallow pinning potentials, whereas the situation is reversed for strong pinning, in excellent accordance with our observation of the crossover to $j_c(B_6) > j_c(B_7)$ at lower temperatures. Note, however, that one caged vortex in a honeycomb lattice restores the ideal hexagonal vortex lattice, while in our quasi-kagom\'e lattice one interstitial vortex is still associated with elastic energy of the vortex lattice. Thus, additional simulations with varying pinning potentials in the quasi-kagom\'e lattice are envisaged.

\section{CONCLUSIONS}

In summary, we have demonstrated the fabrication of periodic arrays of pinning centers in thin YBCO films with ultranarrow lattice spacings down to 70\,nm by irradiation with the defocused beam of a HIM. This technique opens the route to create large arrays of pinning sites with user-defined positions and with a resolution superior to the lithographic techniques. As an example of a complex pinning landscape, the quasi-kagom\'e pinning lattice exhibits an unconventional commensurability effect, when all pins are occupied by vortices and one interstitial vortex is magnetically caged in each void of the lattice. It has been suggested that such caged vortices can be more easily manipulated along predetermined paths \cite{TOGA05} and, hence, our findings can pave the way toward ``Fluxonic'' applications \cite{WAMB99,HAST03,MILO07} of cuprate superconductors.

\section{MATERIALS AND METHODS}

{\bf \sffamily \ \ Superconducting Film Growth.} Thin films of YBa$_2$Cu$_3$O$_{7-\delta}$ are grown epitaxially on (100) MgO single-crystal substrates by pulsed-laser deposition using 248~nm KrF-excimer-laser radiation at a fluence of 3.2~J/cm$^2$. The thicknesses of the films are $t_z = 75 \pm 5$\,nm (sample QK90) and $t_z = 50 \pm 5$~nm (sample QK70) as determined by atomic force microscopy. Bridges with dimensions $240 \times 60\ \mu \mathrm{m}^2$ are lithographically patterned for the electrical transport measurements. Two longitudinal voltage probes with a distance of $100\ \mu \mathrm{m}$ are applied on side arms of the bridges. The as-prepared samples had critical temperatures $T_c \sim 90$~K and transition widths $\Delta T_c \sim 1$~K.

{\bf \sffamily Ion Beam Irradiation.} The prepatterned YBCO microbridges are introduced into the Zeiss Orion NanoFab He ion microscope and with low ion fluence the proper alignment of the sample surface to the focus plane of the ion optics is checked. Controlled defocus is achieved by changing the working distance by $8\,\mu$m from the focus plane, resulting in irradiation spots of about 20~nm diameter. Every point of the lattice, defined by a deflection list loaded into the Nano Patterning and Visualization Engine (NPVE), is irradiated with 30 keV He$^+$ ions with a dwell time of 1.2\,ms (0.83\,ms) and a beam current of 2.0\,pA (2.9\,pA) for sample QK90 (QK70), corresponding to 15000 He$^+$ ions/point. The average fluence hitting the sample's surface in the focus spot of about 20\,nm fwhm is $4.8 \times 10^{15}\,\text{cm}^{-2}$. The irradiated area is approximately $200\,\mu{\rm m}\times 100\,\mu{\rm m}$.

{\bf \sffamily Electrical Transport Measurements.} The electrical measurements are performed in a closed-cycle cryocooler mounted between the pole pieces of an electromagnet. A Cernox resistor \cite{HEIN98} together with a LakeShore 336 temperature controller is used for in-field temperature control to a stability of about 1 mK. The magnetic field, oriented perpendicular to the sample surface, is tuned by a programmable constant current source and measured with a LakeShore 475 gaussmeter with a resolution of 0.1 $\mu$T, a zero offset $< 10 \mu$T, and a reading accuracy $<0.1~\%$. The resistivity measurements are performed with a current of 0.8\,mA, generated by a Keithley 6221 constant-current source in both polarities to exclude thermoelectric signals, and the voltage is measured with a Keithley 2182A nanovoltmeter. The critical current is determined from isothermal current-voltage ($I-V$) measurements with a voltage criterion of 100~nV, corresponding to $10\,\mu$V/cm. The $I-V$ curves do not exhibit features that could be used to discriminate between the  depinning of the vortices trapped in the CDs and the interstitial vortices, respectively.

{\bf \sffamily Numerical Simulations.}
The simulations are performed in a 2D (in the $xy$-plane) simulation cell with periodic boundary conditions that models an infinite superconducting film characterized by the  penetration depth $\lambda$, which is set to $\lambda_{L,ab}(T)$ at finite $T$.
The cell is chosen large enough to avoid finite-size effects. Simulated annealing calculations are performed by numerically integrating the overdamped equations of motion \cite{REIC98b,MISK05,MISK06}:
$
\eta {\rm \bf v}_{i} \ = \ {\rm \bf f}_{i} \ = \ {\rm \bf f}_{i}^{vv} + {\rm \bf f}_{i}^{vp} + {\rm \bf f}_{i}^{T} + {\rm \bf f}_{i}^{d}.
$
Here
${\rm \bf f}_{i}$
is the total force per unit length acting on vortex
$i$,
${\rm \bf f}_{i}^{vv}$
and
${\rm \bf f}_{i}^{vp}$
are the forces due to vortex-vortex and vortex-pin interactions, respectively,
${\rm \bf f}_{i}^{T}$
is the thermal stochastic force,
and
${\rm \bf f}_{i}^{d}$
is the driving force;
$\eta$ is the viscosity, which is set to unity.
The force due to the vortex-vortex interaction is
$
{\rm \bf f}_{i}^{vv} \ = \ \sum_{j}^{N_{v}} \ f_{0} \ K_{1} \!
\left( \mid {\rm \bf r}_{i} - {\rm \bf r}_{j} \mid / \lambda \right)
\hat{\rm \bf r}_{ij},
$
where
$N_{v}$
is the number of vortices,
$K_{1}$
is a modified Bessel function,
$\hat{\rm \bf r}_{ij} = ( {\rm \bf r}_{i} - {\rm \bf r}_{j} )
/ \mid {\rm \bf r}_{i} - {\rm \bf r}_{j} \mid,$
and
$
f_{0} = \Phi_{0}^{2} / 8 \pi^{2} \lambda^{3} .
$
The pinning force is
$
{\rm \bf f}_{i}^{vp} = \sum_{k}^{N_{p}}  f_{p} \cdot \left( \mid {\rm \bf r}_{i} - {\rm \bf r}_{k}^{(p)} \mid / r_{p} \right)
\Theta \!
\left[ \left(
r_{p} - \mid {\rm \bf r}_{i} - {\rm \bf r}_{k}^{(p)} \mid \right)/\lambda
\right]
\hat{\rm \bf r}_{ik}^{(p)},
$
where
$N_{p}$
is the number of pinning sites,
$f_{p}$ (expressed in $f_{0}$)
is the maximum pinning force of each short-range parabolic potential well
located at ${\rm \bf r}_{k}^{(p)}$, $r_{p}$
is the range of the pinning potential,
$\Theta$ is the Heaviside step function,
and $\hat{\rm \bf r}_{ik}^{(p)} = ( {\rm \bf r}_{i} - {\rm \bf r}_{k}^{(p)} )
/ \mid {\rm \bf r}_{i} - {\rm \bf r}_{k}^{(p)} \mid$.
All the lengths (fields) are expressed in units of
$\lambda$ ($\Phi_{0}/\lambda^{2}$).
The ground state of a system of moving vortices is obtained by simulating field-cooled experiments. In this approximation of deep short-range ($\delta$-like) potential wells, the critical current can be defined as follows
\cite{REIC98b,MISK05,MISK06,MISK10,BOTH14}
(giving essentially the same results as those obtained using the threshold criterion in dynamical simulations \cite{MISK05,MISK06}):
$
j_{c}(\Phi) = j_{0} N_{v}^{(p)}(\Phi) / N_{v}(\Phi),
$
where
$j_{0}$
is a constant,
and study the dimensionless value
$J_{c} = j_{c}/j_{0}$.

\section*{ASSOCIATED CONTENT}
{\bf \sffamily Supporting Information}\\
\noindent The Supporting Information is available free of charge on the ACS Publications website at \href{https://doi.org/10.1021/acsanm.9b01006}{DOI: 10.1021/acsanm.9b01006}

\begin{quote}
Critical current and resistance data of a square array of vortex pinning centers fabricated by irradiation in the helium ion microscope (PDF)
\end{quote}

\section*{AUTHOR INFORMATION}
{\bf \sffamily Corresponding Author}\\
*E-mail: wolfgang.lang@univie.ac.at\\
{\bf \sffamily ORCID}\\
Bernd Aichner: 0000-0002-9976-9877\\
Vyachelav R. Misko: 0000-0002-5290-412X\\
Meirzhan Dosmailov: 0000-0003-3672-0619\\
Johannes D. Pedarnig: 0000-0002-7842-3922\\
Franco Nori: 0000-0003-3682-7432\\
Wolfgang Lang: 0000-0001-8722-2674\\
{\bf \sffamily Present Address}\\
M.D.: Al-Farabi Kazakh National University, Almaty, Kazakhstan\\
{\bf \sffamily Notes}\\
The authors declare no competing financial interest.

\section*{ACKNOWLEDGMENTS}

The authors thank John Notte and Jason Huang for enlightening discussions on helium ion microscopy and for initial test irradiations in the HIM at Carl Zeiss Ion Microscopy Innovation Center in Peabody, USA, and Georg Zechner for experimental support.
B.M. acknowledges funding by the German Academic Scholarship Foundation.
V.R.M. and F.N. acknowledge support by the
Research Foundation-Flanders (FWO-Vl) and Japan Society for the Promotion of Science (JSPS) (JSPS-FWO Grant VS.059.18N).
F.N. is supported in part by the
MURI Center for Dynamic Magneto-Optics via the Air Force Office of Scientifc Research (AFOSR) (FA9550-14-1-0040),
Army Research Office (ARO) (Grant 73315PH),
Asian Office of Aerospace Research and Development (AOARD) (Grant FA2386-18-1-4045),
Japan Science and Technology Agency (JST)
(the Q-LEAP program and CREST Grant JPMJCR1676),
Japan Society for the Promotion of Science (JSPS)
(JSPS-RFBR Grant 17-52-50023),
RIKEN-AIST Challenge Research Fund.
Research was conducted within the framework of the COST Action CA16218 (NANOCOHYBRI) of the European Cooperation in Science and Technology.

\section*{REFERENCES}
\vskip -30pt
{\footnotesize \bibliography{HIM_nano_final}}

\end{document}